# Elucidating dislocation core structures in titanium nitride through high-resolution imaging and atomistic simulations

J. Salamania[1*], D.G. Sangiovanni[2], A. Krayich[3], K.M. Calamba Kwick[4], I.C. Schramm[4], L.J.S. Johnson[4], R. Boyd[1], B. Bakhit[5], T.W. Hsu[1], M. Mrovec,[3] L. Rogström[1], F. Tasnádi[2], I.A. Abrikosov[2], and M. Odén[1]

[1]*Nanostructured Materials Division, Department of Physics, Chemistry and Biology (IFM), Linköping University, Linköping, SE-581 83, Sweden*

[2]*Theoretical Physics Division, Department of Physics, Chemistry and Biology (IFM), Linköping University, Linköping, SE-581 83, Sweden*

[3]*The Interdisciplinary Centre for Advanced Materials Simulation (ICAMS), Ruhr-Universität Bochum, D-44780 Bochum, Germany*

[4]*Sandvik Coromant AB, Stockholm, SE-126 79, Sweden*

[5]*Thin Film Physics Division, Department of Physics, Chemistry and Biology (IFM), Linköping University, Linköping, SE-581 83, Sweden*

*Corresponding author: janella.salamania@liu.se

**ABSTRACT**

Although titanium nitride (TiN) is among the most extensively studied and thoroughly characterized thin-film ceramic materials, detailed knowledge of relevant dislocation core structures is lacking. By high-resolution scanning transmission electron microscopy (STEM) of epitaxial single crystal (001)-oriented TiN films, we identify different dislocation types and their core structures. These include, besides the expected primary $a/2\{110\}\langle 1\bar{1}0\rangle$ dislocation, Shockley partial dislocations $a/6\{111\}\langle 11\bar{2}\rangle$ and sessile Lomer edge dislocations $a/2\{100\}\langle 011\rangle$. Density-functional theory and classical interatomic potential simulations complement STEM observations by recovering the atomic structure of the different dislocation types, estimating Peierls stresses, and providing insights on the chemical bonding nature at the core. The generated models of the dislocation cores suggest locally enhanced metal-metal bonding, weakened Ti-N bonds, and N vacancy-pinning that effectively reduces the mobilities of $\{110\}\langle 1\bar{1}0\rangle$ and $\{111\}\langle 11\bar{2}\rangle$ dislocations. Our findings underscore that the presence of different dislocation types and their effects on chemical bonding should be considered in the design and interpretations of nanoscale and macroscopic properties of TiN.

**Nomenclature**

| | |
|---|---|
| TiN | : titanium nitride |
| MgO | : magnesium oxide |
| TMN | : transition metal nitride |
| FCC | : face-centered cubic |
| B1 | : rock-salt structure |
| UHV | : ultra-high vacuum |
| TEM | : transmission electron microscopy |
| STEM | : scanning transmission electron microscopy |
| HAADF | : high-angle annular dark-field |
| DFT | : density-functional theory |
| MEAM | : modified embedded atom method |
| XRD | : X-ray diffraction |
| ToF-ERDA | : time-of-flight elastic recoil detection analysis |
| BF | : bright field |
| FIB | : focused ion beam |
| SEM | : scanning electron microscope |
| FFT | : fast Fourier transform |
| DOS | : density of states |
| FWHM | : full width at half maximum |
| GSFE | : generalized stacking fault energy |
| OP | : overlap-population |

**1. Introduction**

Crystallographic dislocations are ubiquitous one-dimensional line defects [1, 2] that profoundly influence material's mechanical [3, 4], chemical [5], optical [6], electronic [7, 8], and magnetic [9] properties. The elastic distortions and strain fields generated by the dislocation core locally modify the atomic and electronic structure [10]. Dislocations may act as sinks for impurities and vacancies [7, 11], affect local composition that trigger phase transformations [12], and act as pathways for enhanced mass transport (pipe-diffusion) [13, 14] thereby affecting macroscopically observed properties. More importantly, the nucleation, motion, multiplication, and interaction of dislocations [15] govern plastic deformation and is determined by operative slip systems [1, 2]. Such favored slip systems and other dislocation-related mechanisms are intimately associated to the core structure of its dislocations. In many aspects, a thorough characterization of dislocation cores at the atomic- and electronic-structure level is of the utmost importance for understanding various material's properties.

Titanium nitride (TiN) is a well-studied transition metal nitride (TMN) and is used in hard coating [16-19], microelectronic [20], optical [21], and plasmonic material applications [22]. It is thermally and electrically conductive, wear-resistant, and has a high thermal stability owing to its rock-salt (B1) structure that consists of bonds with mixed ionic, metallic, and covalent character [23-25]. Because of this, the dislocation structures and slip in TiN are affected by the directional nature of the bonds near the core [26-28]. Complementary electronic-structure analyses are therefore necessary to probe the effect of core-induced lattice distortion on the local bonding character.

Dislocations are well-studied defects in metals and semiconductors [29] but only few reports have tried to elucidate the dislocation core structures in ceramics, specifically transition metal nitrides, at the atomic scale [3, 7, 30]. Previous studies on dislocations in TiN were based on *ab initio* calculations, nanoindentation of single crystal films, or microscopy observations of polycrystalline samples [19]. *Ab initio* evaluations of Peierls stresses by Yadav *et al.* [23] have shown that the preferred slip planes for TiN are {110}, followed by {111} and then {100}. Their *ab initio* calculations also suggest an energetically-favored dissociation of the a/2 {111}$\langle 1\bar{1}0 \rangle$ dislocation into two a/6$\langle 11\bar{2} \rangle$ Shockley partial dislocations [23], which is a mechanism typical for face-centered cubic (FCC) crystals [1, 31]. Experimentally, it has been observed that the $\{111\}\langle 1\bar{1}0\rangle$ [16] and $\{110\}\langle 1\bar{1}0\rangle$ [24] slip systems are activated during room-temperature [32-34] and high-temperature [35] nanoindentation of thin lamellas along the $\langle 111\rangle$ and $\langle 100\rangle$ directions. The observation of the $\{111\}\langle 1\bar{1}0\rangle$ slip, however, was through nucleation of Shockley partials during indentation in the $[\bar{1}11]$ direction [16]. TEM observations of edge dislocations were also reported for polycrystalline TiN, including the relation between dislocation density and grain sizes [36].

While perfect $\{110\}\langle 1\bar{1}0\rangle$ edge dislocations and Shockley partial dislocations have been observed experimentally through controlled deformation of TiN, other extended crystallographic defects, such as stacking faults and sessile Lomer dislocations, are less well studied or not even reported in TiN or other B1 carbides and nitrides. Thus, knowledge of the atomic-scale structure and energetics of these defects is mostly lacking. Accurate identification and characterization of different dislocation core structures is important for understanding their effect on macroscopic properties. For example, the density and mobility of pre-existing dislocations around crack tips and grain boundaries largely govern plasticity and thus affect the toughness of ceramics [4, 18, 33]. Moreover, studies of preferred dislocation core structures show that atomic-scale distortions at the core can help to explain a material's thermal-electric breakdown [7]. Further, understanding of the preferred dislocation structures and their cores can potentially be used for improving ceramic properties and performance through dislocation engineering and toughening mechanisms [4, 37, 38].

Detailed atomic-level characterization of dislocations in TiN by transmission electron microscopy is typically hindered by high densities of defects, such as interstitials, vacancies, dislocations, and internal interfaces present at phase and grain boundaries and macroparticles present in coatings commonly grown for industrial applications [39-42]. Specimens with high crystalline quality and low density of dislocations are thus desirable to resolve dislocation core structures with atomic resolution by transmission electron microscopy (TEM) or scanning transmission electron microscopy (STEM). This can be challenging since dislocations are readily introduced into crystals during growth [1]. In other material classes, this challenge has been typically addressed by utilizing a bicrystal technique [7, 8, 43], where two pristine single crystal blocks are joined together by diffusion bonding at high temperatures. For nitride films, we address the same challenge by synthesizing single crystal films with high crystallinity through ultrahigh vacuum (UHV) magnetron sputtering methods [39, 40]. The growth of heteroepitaxial single crystal films of TiN on magnesium oxide (MgO) is therefore ideal because they are free of grain boundaries and have sufficiently low defect density to make them suitable for detailed microstructural STEM characterizations.

Here, we report the atomic structure and characterize the dislocation cores in TiN on the known primary $\{110\}\langle 1\bar{1}0\rangle$ slip system as well as on the non-primary $\{111\}\langle 1\bar{1}0\rangle$ and $\{100\}\langle 011\rangle$ slip systems. This is done by, first, growing high-quality single crystal TiN films using UHV magnetron sputtering (Section 2.1); then, characterizing the film and imaging dislocations with high-resolution using scanning transmission electron microscopy (STEM), Section 2.2. We complement our observations by using density-functional theory (DFT) and modified embedded atom method (MEAM) potential simulations to gain insights into the nature of bonding at the dislocation cores (Section 2.3). The findings are presented in Section 3 and the implications discussed in Section 4.

**2. Methods**

*2.1 Film growth*

Single crystal TiN thin films were deposited using a Mantis ultrahigh vacuum DC magnetron sputtering system with a base pressure of less than 4 x $10^{-10}$ Torr (5 x $10^{-8}$ Pa) following the procedure outlined elsewhere [39]. The substrates (10x10x0.5 mm$^3$ single crystal MgO(001) of >99.5 % purity) were subjected to systematic chemical pre-cleaning [44], and a 30 min vacuum anneal at 800°C in the deposition chamber just before starting film growth. The TiN films were reactively sputter-deposited from a single element Ti target (99.9% purity) in a 20 sccm Ar to 4 sccm $N_2$ gas mixture resulting in a working pressure of 0.3 Pa. Before deposition, the Ti target was sputter-cleaned at low power in Ar plasma with closed shutters, protecting the substrate. During deposition, the substrate temperature was kept at 800°C while the substrate holder was rotated at 10 rpm. The Ti target was supplied with a fixed 0.4 A and varying voltage up to 600 V resulting to a deposition rate of ~4.7 nm/min. After deposition, the samples were allowed to cool slowly to less than 50°C before transferring to a load-lock chamber and venting the system.

*2.2 Characterizations and imaging*

X-ray diffractograms (XRD) of the thin films were acquired using a PANalytical Empyrean diffractometer for crystal structure and orientation analysis. A channel-cut 2-bounce Ge(220) hybrid monochromator was used as the primary optics for the measurements while a 3-bounce Ge(220) symmetrical analyzer or a parallel plate collimator was used as the diffracted beam optics. High-resolution XRD measurements were obtained with time step of 3.25 seconds and step size of 0.002°. All measurements were performed using Cu K$\alpha$ ($\lambda$ = 1.54187 Å) radiation.

Elemental compositions of the thin films were measured by time-of-flight elastic recoil detection analysis (ToF-ERDA) using a 36 MeV $^{127}I^{8+}$ beam in a 5 MV 15SDH-2 tandem accelerator. More details about the data acquisition and analysis can be found in [45].

Sample surface morphologies were imaged using a Zeiss LEO 1550 Gemini scanning electron microscope at 3 kV acceleration voltage. Cross-section TEM lamellas were prepared by an *in-situ* lift-out technique in two focused ion beam (FIB) - scanning electron microscope (SEM) dual beam systems (FEI Helios Nanolab 650 and Zeiss Neon 40). A protective platinum layer was first deposited by both electron beam and ion beam prior to FIB milling to prevent beam damage from Ga$^+$ ions. To prepare plan-view lamellas, the samples were first thinned to ~75 μm thickness from the substrate side by a combined mechanical cutting, cleaving, and polishing process before FIB thinning. Final milling steps of 5 kV/15 pA and 2 kV/9pA were used to minimize any amorphous layers generated using high voltage thinning. The TEM lamellas were prepared in such a way that the sample can be easily oriented in the $\langle 100 \rangle$ and $\langle 110 \rangle$ zone axes during imaging and were fixed to a Cu grid.

The bright field (BF) and high-angle annular dark field (HAADF) STEM images were acquired from an aberration-corrected FEI Titan[3] transmission electron microscope operated at 300 kV. The HAADF-STEM images were recorded using an angular detection range of 50-200 mrad. HAADF-STEM high-resolution imaging was done using a 21.4 mrad semi angle probe and 54 pA beam current.

Image analysis using inverse fast Fourier transform (FFT) and Wiener filtering was performed using Gatan's Digital Micrograph software. The presented dislocations and their corresponding Burgers vectors were identified via the finish-to-start right-hand convention. Denoising of high-resolution STEM images were performed using the AtomSegNet software implemented by Lin *et al.* [46]. The multislice simulation approach, implemented in the Dr. Probe software [47], was used to simulate HAADF-STEM images from generated atomic models (discussed below) using similar parameters as applied during actual acquisition of images.

*2.3 Atomistic simulations*

Calculations of dislocation core structures and their properties were based on quantum mechanics description in the framework of density functional theory (DFT) and on classical simulations using the modified embedded atom method (MEAM) potential, as parameterized in Ref. [48]. The core structures of $\{110\}\langle 1\bar{1}0\rangle$ edge dislocations, $\{111\}\langle 11\bar{2}\rangle$ Shockley partial dislocations, and $\{100\}\langle 011\rangle$ Lomer edge dislocations were initially generated using the Atomsk code [49]. Thereafter, the atomic positions were relaxed using the MEAM potential implemented in the LAMMPS software [50]. The reliability of our MEAM model has been proven by comparison with microscopy observations of TiN-based materials subject to load as well as DFT and *ab initio* molecular dynamics simulations of mechanical and thermodynamic properties [51-53]. The dislocation core energies were assessed using different supercell configurations (details in Supplementary Material (SM)). For Peierls stress $\tau_P$ evaluations, we employed supercells with ≥ 5000 atoms which contain an isolated core. The supercells are periodic only parallel to the dislocation line. Shear deformation was incremented up to dislocation glide. The Peierls stress was determined as the shear stress required to activate dislocation motion.

DFT calculations were carried out using the VASP code implemented with the projector augmented wave method [54, 55]. Effects of exchange and correlation on one-electron potentials and total energies were evaluated using the generalized gradient approximation with the parametrization by Perdew, Burke, and Ernzerhof [56]. We employed a cutoff energy of 500 eV and Γ-point sampling of the reciprocal space. The dislocation cores obtained by MEAM structural optimization were used as input into DFT calculations. The supercells contained ≈1200 atoms and were periodic only along the dislocation line. Relaxation of atomic positions around the dislocation core (see below) were iterated up to an accuracy of $10^{-5}$ eV/supercell, with forces on atoms smaller than 0.01 eV/Å. The electronic density of states (DOS) and electron-transfer maps were calculated for the relaxed configurations. The latter were obtained by subtracting the overlapping electron density of non-interacting atoms from the self-consistent charge density of the relaxed supercells. The effect induced by local lattice distortions on $Ti_{core}$–$N_{core}$ bond strengths has been assessed by partitioning the wavefunctions into two-center molecular orbitals [57]. Hence, we employed a DFT-based approach to extract the overlap population (OP) of Ti–N bonds in bulk regions and at the dislocation core. The OP is a projection of the DOS into bonding (OP > 0) or antibonding (OP < 0) interactions between different pairs of atomic orbitals.

**2. Results**

*3.1 Characterization of as-deposited films*

Figure 1 shows representative ω-2θ diffractograms of the sputter-deposited TiN (001) films on MgO (001) substrates. Only the TiN 200 (~42.5°) peak was observed from the films besides their corresponding MgO 200 substrate peak. The inset high-resolution ω-2θ scan reveals visible finite-thickness interference fringes (Laue oscillations) indicating high structural epitaxial quality of the films, smooth surfaces, and uniform substrate to film interfaces [20, 58]. The full width at half maximum (FWHM) of the high-resolution ω-2θ TiN 200 film peak is 0.023° compared to 0.002° of the MgO 200 substrate peak. The inserted rocking curve shows that the TiN film is highly orientated with a narrow

FWHM of less than 216 arcsecs or 0.06° of the 200 film peak. The x-ray diffractometry results show that the deposited layers are heteroepitaxial, in agreement with previous results using similar deposition conditions [39]. The dislocation density of the film estimated from the rocking curve FWHM is in the order of $10^8$ cm$^{-2}$ by using the Hirsch model [59] and Burgers vector a/2 ⟨110⟩. For comparison, previous experimental investigations reported dislocation densities < $10^7$/cm$^2$ for (001)-oriented and < $10^9$ cm$^{-2}$ for (111)-oriented single-crystal TiN films [60]. The film surfaces were smooth and featureless consistent with the previously observed epitaxial TiN films grown at high deposition temperatures [20]. ToF-ERDA measurements reveal that the deposited TiN films are slightly substoichiometric with a N/Ti content ratio of 0.97 ± 0.02 suggesting the presence of nitrogen vacancies. The measured O and C contents in the films are 1.0 ± 0.1% and 0.6 ± 0.1%, respectively, which are at the detection accuracy limit of ToF-ERDA.

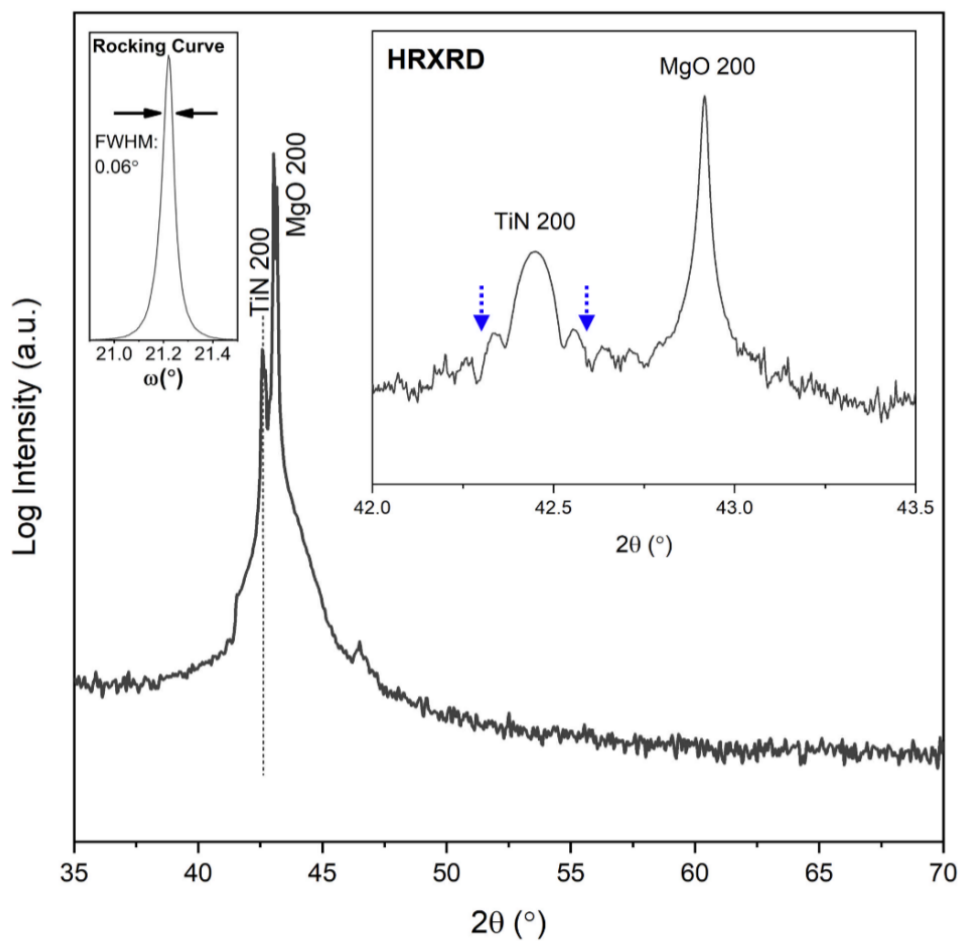

Figure 1. X-ray ω-2θ diffractogram of the sputter-deposited TiN (001) films grown on MgO (001) substrates. The insets show a ω-rocking curve of the TiN 200 diffraction line and a high-resolution ω-2θ diffractogram (HRXRD) displaying interference fringes pointed by the blue arrows.

Cross-sectional STEM imaging reveals a homogeneous thin film of ~140 nm thickness free of grain boundaries, while FFT analysis confirms the cube-on-cube epitaxial relationship of the film to the substrate (see Fig. S1). The HAADF-STEM images show that the film is heteroepitaxial at the film-substrate interface with no observable misfit dislocations but with slight local lattice misorientation, as evidenced by the varying contrast seen as faint columns along the growth direction of the film. Such variation in contrast could also be affected by the HAADF-STEM intensity changes caused by strain relaxations on the surface of the lamella [61].

*3.2 Observed dislocations and their core structures*

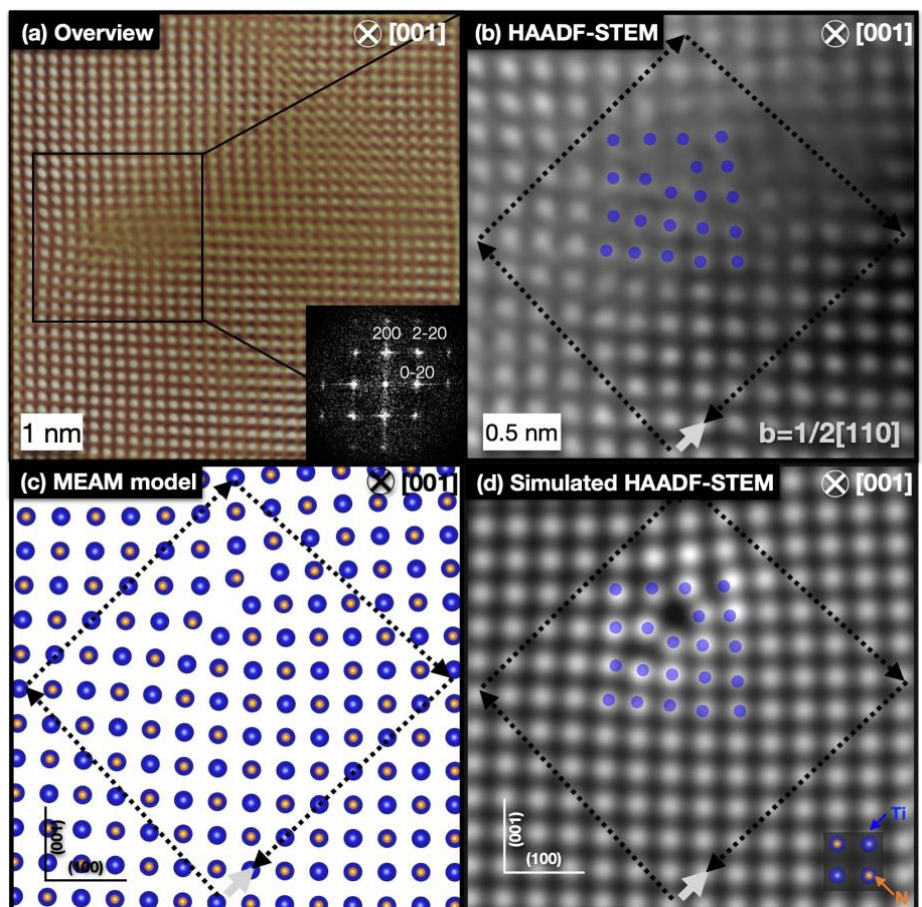

Figure 2. (a) Overview cross-sectional HAADF-STEM image of a region containing an end-on full edge dislocation with projected Burgers vector of a/2[110] in TiN viewed from the [001] direction. To facilitate visualization, the inverse FFT was superimposed on the image with 25% transparency. (b) Magnified view of region containing the edge dislocation. The blue dots aligned with the atomic columns in the experimental image are translated directly to the simulated STEM image in (d) and only serve as guide for the reader. (c) MEAM model of an edge dislocation with b=a/2[110] viewed from [001] zone axis. (d) Simulated HAADF-STEM image based on the dislocation core model in (c).

Figures 2a and 2b show high-resolution plan-view HAADF-STEM images of an end-on full edge dislocation with burgers vector a/2[110] viewed along the [001] zone axis. The presence of the edge a/2 {110}⟨1$\bar{1}$0⟩ dislocation type shown in Figure 2 is expected since {110}⟨1$\bar{1}$0⟩ is commonly reported as the most active slip system in B1 nitrides and carbides at room temperature [16, 24, 60, 62] and predicted to have the lowest Peierls stress [23]. The revealed core geometry is closely reproduced in the simulated HAADF-STEM image (Fig. 2d) of the theoretically predicted atomic structure obtained by classic simulations with MEAM force field (Fig. 2c). The atom positions recorded in the HAADF-STEM image are marked by blue dots (Fig. 2b), which are then directly superimposed on the simulated HAADF-STEM image (Fig. 2d) to facilitate comparison and guide the reader. Unclosed Burgers circuits are also drawn for guidance. The contrast around the dislocation core in the HAADF-STEM image is blurred due to the strain fields induced by the distorted lattice as well as the imperfect alignment of the core to the electron beam. The good qualitative match between experimentally observed and simulated dislocation cores validates our MEAM-predicted model in terms of accurately describing the {110}⟨1$\bar{1}$0⟩ core configuration.

The local electronic structure at the core of this type of dislocation was calculated from first principles in the framework of DFT (Fig. 3). The initial configuration of our DFT supercell model is taken from the MEAM-relaxed atomic positions. The atomic coordinates within the red-dashed region in Fig. 3a are relaxed by DFT, while all other atoms are kept fixed throughout the calculation. The core structure predicted from first principles maintains similar bond distances and angles in comparison with results of the classical simulations. Electron transfer maps around the core are calculated on both a (001) plane perpendicular to the dislocation line, Fig. 3c, and on the (110) plane of symmetry of the dislocation core, Fig. 3e. For reference, Fig. 3b and 3d show electron transfer maps calculated for "bulk" domains of defect-free (B1) TiN. In "bulk" domains, Ti and N atoms exhibit spherical electron distributions – with Ti sites characterized by electron depletion (red) and N sites by electron accumulation (blue) – which is typical for ionic-like bonding, similar to what previously reported for defect-free TiN [28, 63, 64].

The electron transfer map at the dislocation core shown in Fig. 3b suggests enhanced *d-d* bonding between some Ti neighboring atoms located near the core. The effect is indicated by electron accumulation—blue color between, e.g., $Ti_1$ and $Ti_2$, and $Ti_3$ and $Ti_4$. In most cases, however, the Ti – Ti d-states around the core, and on the (001) plane, are essentially non-bonding. This is understood by noting electron depletion between, e.g., $Ti_5$ and $Ti_6$, which is similar to the configuration in the bulk domain (yellow color between neighbor $Ti_b$ atoms in insets of Fig. 3b, 3d). In contrast, a cut of the charge density along the mirror-plane of the core – marked as a solid green line in Fig. 3a – reveals electron accumulation parallel to the dislocation line (blue color between $Ti_1$-N-$Ti'_1$ and $Ti_2$-N-$Ti'_2$ lattice rows in Fig. 3e). This is reflected by a locally enhanced population of d electronic states near the Fermi level, as discussed below. The origin of the one-dimensional localization of d electrons can be interpreted as follows: the slight undulation of the lattice rows along [001] around the core yields shorter second-neighbor Ti-Ti distances along $[\bar{1}10]$, which enhances *d-d* orbital overlapping compared to defect-free bulk regions [64]. In addition, the zig-zag conformation of Ti-N-Ti [001]-rows only partially screens fourth-neighbor Ti-Ti interactions along [001], similar to metal-metal interactions across anion vacancy sites in understoichiometric B1 nitrides [65]. In general, the electron-transfer maps of Fig. 3 provide an example of how the electronic structure and bonding nature at the core of the dislocation are substantially altered in comparison to defect-free regions.

The films were also imaged along the [011] zone axis to view dislocations in the non-primary {111}⟨110⟩ slip system. Figure 4 shows a representative HAADF-STEM image of a Shockley partial dislocation corresponding to a Burgers vector of $a/6[1\bar{1}2]$ by applying Burgers circuit analysis. An offset (marked with a dashed white line) between $(\bar{1}1\bar{1})$ lattice planes was observed along the unclosed Burgers circuit. A simulated HAADF-STEM image from the MEAM model in Fig. 5 below is placed as an inset in Fig. 4c and shows good match with the experimentally observed partial dislocation core.

Previous *in-situ* TEM imaging studies in the [011] zone axis observed the nucleation of a partial dislocation on {111} planes by selectively indenting a thin lamella along the ⟨111⟩ direction [16]. They also estimated a critical shear stress of 13.8 GPa for nucleating a partial dislocation in TiN [16]. To compare, nanoindentation hardness measurements of (001)-oriented TiN films were reported to be in the range 17.3 to 22.1 GPa, which corresponds to maximum 2.8–3.6 GPa shear stresses induced in the film [60]. The latter values – comparable to the Peierls stresses estimated by DFT for glide on {111} and {110} planes [23] – suggest that the stress induced in the films during hardness measurements are too low to nucleate partial dislocations, but likely sufficient to produce glide of pre-existing {111}⟨1$\bar{1}$0⟩ and {110}⟨1$\bar{1}$0⟩ dislocations. Note, however, that the presence of lattice vacancies may significantly alter the mobility of edge dislocations (see below).

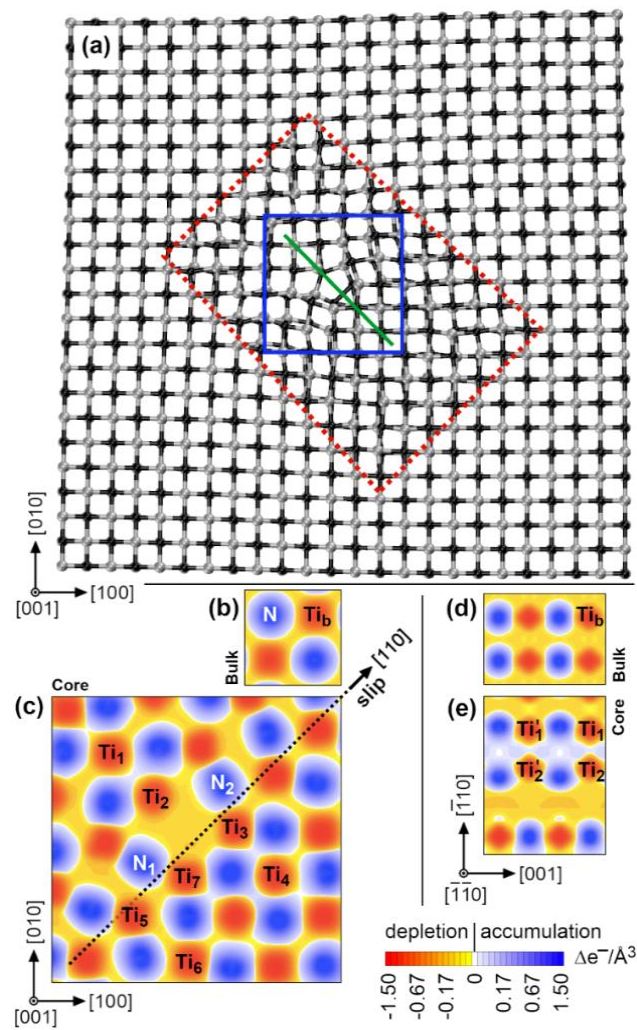

Figure 3. (a) {110}⟨1$\bar{1}$0⟩ edge dislocation relaxed by DFT. Atomic relaxation is performed for atoms within the red rectangle in (a). The cell is periodic along the dislocation line, while vacuum separates supercell replicas along the [100] and [010] crystallographic directions. Ti and N atoms are indicated by silver and black spheres, respectively. The "core" charge density map in (c) corresponds to the area marked by the solid-blue square in (a). The plane of the "core" charge density map shown in panel (e) is indicated by a solid green line in (a). Panels (b) and (d) show electron density maps calculated in a "bulk" domain, i.e., away from the dislocation core.

As described above, the presence of {110}⟨1̄10⟩ and {111}⟨1̄10⟩ edge dislocations in our film grown under conditions chosen to minimize the number of dislocations shows that there are regions where plastic deformation can propagate without the need for homogenous dislocation nucleation. Considering the stochastic nature of dislocation nucleation and local stress inhomogeneities in epitaxial thin films, the presence of these types of dislocations scattered in as-deposited films suggests a glide-dominated rather than dislocation nucleation-dominated plastic deformation during nanoindentation, which is magnified for films grown under conditions that generates more dislocations.

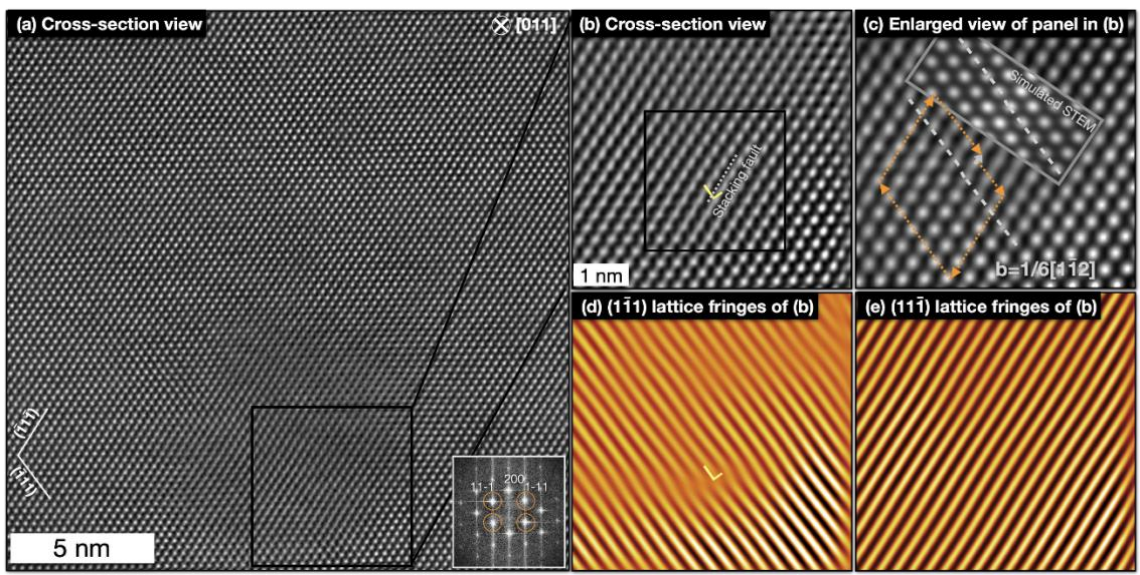

Figure 4. Cross-sectional HAADF-STEM images where (a) and (b) are views from the [011] zone axis showing a Shockley partial dislocation. (c) Magnified view of the dislocation with corresponding Burgers circuit. The inset shows the simulated HAADF-STEM image. (d), (e) Lattice fringes of image (b) obtained by FFT masks of the (1̄11) and (111̄) planes. The yellow "L" marks the location of the Shockley partial dislocation while the white broken lines mark the stacking faults.

The atomistic and electronic structure of Shockley partials in TiN were analyzed via classical and first-principles simulations (MEAM and DFT), as shown in Fig. 5. A TiN supercell that contains two Shockley partials was first relaxed by classical calculations. A portion of the relaxed supercell, visualized in Fig. 5a, served as input for the DFT calculations. Atoms enclosed within the red-dashed line in Fig. 5a were relaxed, while all other atoms were kept fixed at their original sites. As for the case of {110}⟨1̄10⟩ edge dislocations, see close recovery of the lattice positions in the simulated STEM image insert in Fig 4c. Fig. 5b illustrates a [111]-view of the dislocation core. It displays a section of the supercell, which includes four (111) atomic layers, to facilitate visualization of a stacking fault ribbon laterally bounded by Shockley partials. The stacking fault (see middle part of Fig. 5b) is characterized by local *a-b-a-b* stacking sequence, with Ti nearly atop other Ti and N nearly atop other N atoms. The width of the defect is estimated to be around 1 nm. A relatively narrow stacking fault region is due to high intrinsic stacking fault energy, which is estimated to be 0.95 J/m$^2$ by DFT calculations [66] and 0.72 J/m$^2$ by our classical model calculations (see generalized stacking fault energy surface in Fig. S2 of the SM). For comparison, the stacking fault energies of FCC elemental metals are in a range between 0.01 and 0.35 J/m$^2$ [29]. Moving laterally away from the faulted area, along [1$\bar{1}$0], the material recovers the *a-b-c* stacking sequence of a B1 structure (Fig. 5b).

A [1$\bar{1}$0] view of the stacking fault is provided in Fig. 5c, where each (111) atomic layer is labeled with a stacking position. The figure shows a (1$\bar{1}$0) atomic layer in correspondence of the vertical green line in Fig. 5a. Ti and N atoms within the stacking fault are vertically aligned along the [111] direction. The electron transfer map calculated on this atomic plane suggests enhanced bonding among Ti atoms on different (111) layers of the stacking fault, as indicated by electron accumulation between sites labelled as "Ti" in Fig. 5d. Away from the stacking fault, the Ti-Ti interactions are essentially nonbonding. We note that for B1 transition-metal carbides, the increase in dislocation activity on {111} planes observed at moderate or elevated temperatures has been suggested to stem from enhanced metal-metal bonding [62]. Finite-temperature calculations would be required to verify this hypothesis.

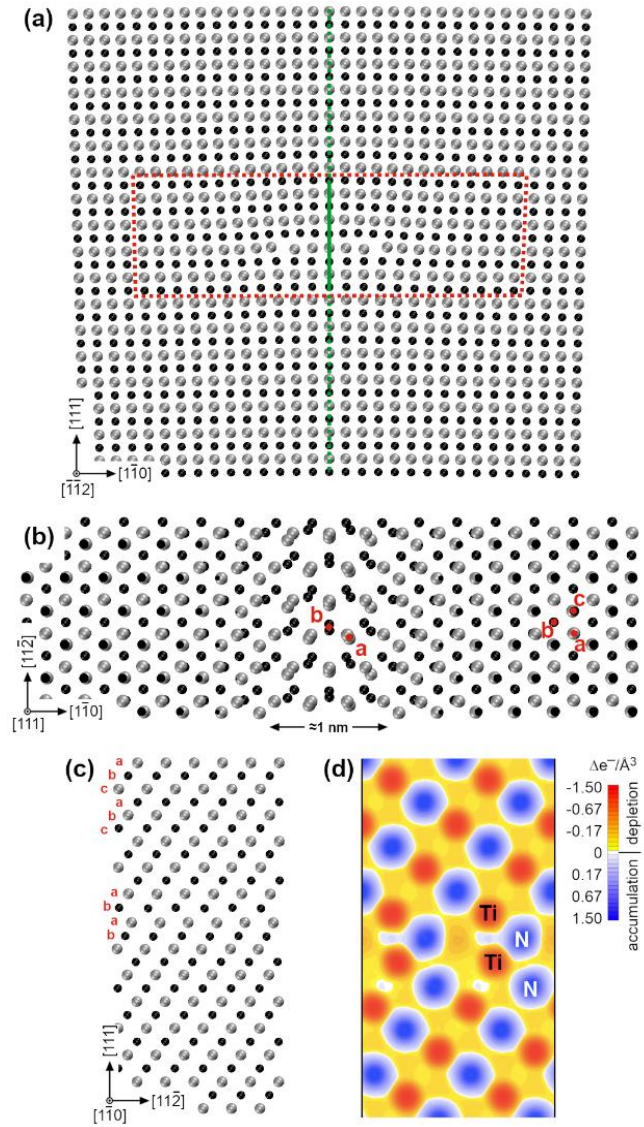

Fig. 5. DFT results for a supercell containing two Shockley partial dislocations. The red-dashed rectangular region in (a) encloses atomic positions relaxed by DFT. Panels (a), (b), and (c) show views from [11$\bar{2}$], [111], and [1$\bar{1}$0] directions, respectively. N and Ti atoms are indicated by black and silver spheres. The supercell is periodic along the dislocation line [11$\bar{2}$]. Vacuum regions separate supercell replicas along [1$\bar{1}$0] and [111] directions. (d) Electron transfer map on a (1$\bar{1}$0) plane that crosses the stacking fault, as indicated by green line in (a).

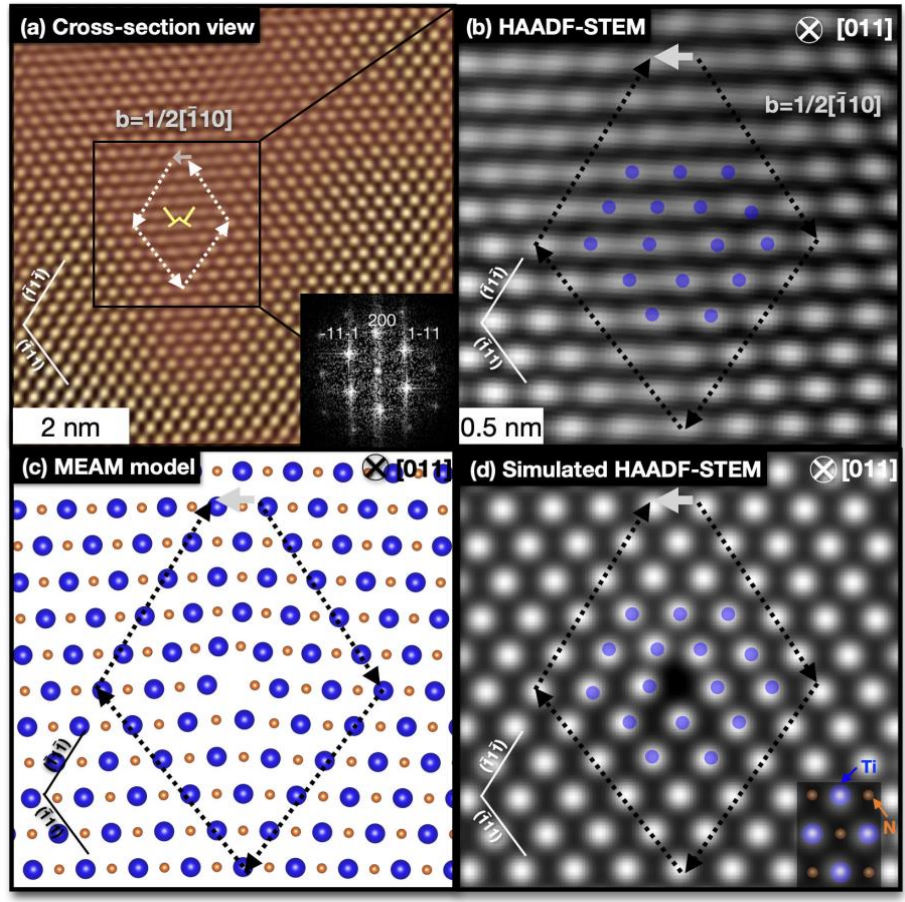

Figure 6. (a) Overview HAADF-STEM image and its corresponding inverse FFT superimposed with 25% transparency. The yellow "⊥"s indicate the locations of the edge dislocations. (b) Magnified view of the region in (a) containing a Lomer dislocation. The blue dots aligned with the atomic columns in the experimental image are translated directly to the simulated HAADF-STEM image in (d) and serve as guide for the reader. (c) MEAM-predicted model of Lomer edge dislocation with b=a/2[110] gliding on (001) plane viewed from [011] zone axis. (d) Simulated HAADF-STEM image based on the model in (c).

Figure 6 presents high magnification HAADF-STEM images and their corresponding inverse FFT showing the presence of two extra half planes on intersecting $(1\bar{1}1)$ and $(11\bar{1})$ planes. This represents a Lomer dislocation (LD) [67, 68] configuration and it is characterized by an edge dislocation with a Burgers vector a/2[011], with slip on (100) planes. The core geometry is in accordance with the structure predicted by classical simulations (Fig. 6c) used to generate the corresponding simulated HAADF-STEM image (Fig. 6d). Like the edge dislocation in Fig. 2, the blurred contrast around the dislocation core here is due to a combination of the accompanying strain fields and projection phenomenon caused by slight misalignment of the electron beam to the dislocation line.

The Lomer defects observed here are sessile in nature [1] and known to induce work-hardening in materials by reaction and locking of sessile {111}⟨1$\bar{1}$0⟩ dislocations [1]. They are found in FCC metals, especially those with large stacking fault energies [69, 70], and FCC high entropy metal alloys [71]. Such dislocations have also been observed as a result of nucleation at an internal interface, accommodating increasing misfit, increasing thickness during growth, and also strain relief during annealing [72]. It is also suggested that LDs can form due to the intersection of gliding dislocations during growth [68, 72]. Dislocations on two intersecting slip planes can attract and trap one another to form LDs producing a configuration with lower core energy [67].

## 3. Discussion

The shortest possible Burgers vector in FCC (B1) structured materials is a/2⟨110⟩ and is therefore expected to be the easiest and most facile glide direction [1, 2]. However, the slip planes of B1 ceramics can be {110}, {111}, or {100} [16, 23]. Depending on the chemical composition, loading direction, and temperature, slip on different planes may be observed [16].

In B1 crystals characterized by ionic and covalent bonding, including TiN [16, 24], a charged-balanced and non-close-packed {110}⟨1$\bar{1}$0⟩ system has been recognized as the primary slip system [73]. In the case of MgO, {001}⟨011⟩ slip is also known to be operative [74], but requires higher activation stresses. There are six distinct {110}⟨1$\bar{1}$0⟩ slip systems. However, the fact that only two {110}⟨1$\bar{1}$0⟩ systems are independent [75] may explain the typical brittleness of B1-structured ceramics. Conversely, in FCC metals, where charge-balance is not important, the close-packed {111}⟨1$\bar{1}$0⟩ system is the predominant slip system [1]. Of the twelve distinct {111}⟨1$\bar{1}$0⟩ slip systems, five are independent and responsible for the characteristic ductility of FCC metals [75].

Although lattice slip on {111} planes is not often observed in B1-structured ceramics, an increase in temperature or change in metal composition can promote its activation. For example, in TiC – also a B1 ceramic like TiN – slip on {110}⟨1$\bar{1}$0⟩ was reported at room temperature, whereas {111}⟨1$\bar{1}$0⟩ slip becomes operative at high temperatures [62]. The temperature-induced change in primary slip system

– from $\{110\}\langle 1\bar{1}0\rangle$ to $\{111\}\langle 1\bar{1}0\rangle$ – is acknowledged as a key mechanism for brittle-to-ductile transitions in group IV carbides [76]. For binary group V carbides, the close-packed $\{111\}\langle 1\bar{1}0\rangle$ system is suggested to be active even at room temperature [62]. The differences in bonding character at the dislocation core may help explain why different B1 transition metal nitrides, carbides, and oxides—despite sharing the same lattice structures—have different preferred slip planes [7, 23, 27, 62]. The presence of partial dislocations and sessile edge dislocations (besides the expected $\{110\}\langle 1\bar{1}0\rangle$ edge dislocations) in TiN calls for complementary investigations of the properties of the different dislocation cores and associated slip systems.

Although dislocations are non-equilibrium defects, the core energy is indicative of the likelihood of observing a given dislocation structure. We carried out classical calculations of dislocation line energies $\Gamma$ – which encompass both the core energy ($E_{core}$) and elastic energy ($E_{ela}$) produced by the dislocation strain field – for $\{110\}\langle 1\bar{1}0\rangle$, $\{111\}\langle 1\bar{1}0\rangle$, and $\{001\}\langle 1\bar{1}0\rangle$ edge dislocations using different simulation setups. Computational details and results are given in the SM. There is no unique way to partition $\Gamma$ into $E_{core}$ and $E_{ela}$ contributions. Moreover, because the investigated dislocation cores glide on different slip planes (*i.e.*, produce different lattice strain fields), the relative differences in core energies cannot be unambiguously determined. Nevertheless, for core radii set equal to $|\mathbf{b}|$, our results indicate that $\{110\}\langle 1\bar{1}0\rangle$, $\{111\}\langle 1\bar{1}0\rangle$, and $\{001\}\langle 1\bar{1}0\rangle$ edge dislocations have similar $E_{core}$ values (between 1.5 and 2.0 eV/Å, see Fig. S3). For comparison, $E_{core}$ determined for edge dislocations in pure FCC Al is approximately 10 times smaller [77]. The relatively small differences in core energies estimated by classical calculations (Fig. S3) are consistent with the coexistence of different edge dislocations in our TiN samples.

The mobility of edge dislocations in stoichiometric TiN and understoichiometric TiN$_x$ (x = 0.97 as in our film) was assessed via classical calculations of Peierls stresses $\tau_P$, which are listed in Table 1. The results obtained for $\{110\}\langle 1\bar{1}0\rangle$ and $\{001\}\langle 1\bar{1}0\rangle$ dislocations in stoichiometric TiN ($\tau_P^{\{110\}}$ = 0.64±0.03 and $\tau_P^{\{001\}}$ = 8.4±0.2 GPa) are in qualitative agreement with DFT values [23] and confirm the relatively high mobility of $\{110\}\langle 1\bar{1}0\rangle$ edge dislocations and sessile nature of Lomer $\{100\}\langle 011\rangle$ dislocations.

However, our results indicate a surprisingly high mobility of Shockley partials. The estimated Peierls stress, $\tau_P^{\{111\}} \approx 0.2$ GPa, is one order of magnitude smaller than DFT predictions ($\tau_P^{\{111\}} \approx 2.7–3.2$ GPa [23]). Such large discrepancy can be due to the limited accuracy of the MEAM potential but can also arise from the simulation setup used in *ab initio* calculations. The 3D-periodic simulation cell employed in DFT [23] probably produces overestimated $\tau_P^{\{111\}}$; considering a stacking fault width of ≈1 nm, the distance between Shockley partials on neighboring cells is insufficient to consider individual core regions as non-overlapping. To support the reliability of our model, we show that the generalized stacking fault energy (GSFE) surface and the stress-strain curves on $\{111\}\langle 1\bar{1}0\rangle$ and $\{111\}\langle 11\bar{2}\rangle$ slip systems obtained by classical calculations compare well with *ab initio* results (see Fig. S2 and detailed discussion in SM). Hence, it is plausible to assume that $\{111\}\langle 1\bar{1}0\rangle$ Shockley partials may easily propagate in TiN domains with ideal 1:1 N/Ti compositional ratio (see supplementary video 1). As showed below, the presence of anion vacancies considerably changes this scenario.

|  | $\{110\}\langle 1\bar{1}0\rangle$ | $\{111\}\langle 1\bar{1}0\rangle$ | $\{001\}\langle 1\bar{1}0\rangle$ |
|---|---|---|---|
| $\tau_P$ (GPa) | 0.64±0.03 (**2.8±0.3**) | ≈0.2 (**6.1±0.1**) | 8.4±0.2 |

Table 1. MEAM-estimated Peierls stresses $\tau_P$ of dislocation structures observed in single-crystal stoichiometric TiN samples. Results in parenthesis are obtained for TiN$_{0.97}$: N vacancies increase the stress required for activation of dislocation motion. In particular, glide on {111} planes become considerably more difficult. Note that a single vacancy is sufficient to obstruct dislocation motion, i.e., the $\tau_P$ values in bold are of practical relevance.

To probe the effects of N deficiency on dislocation activity, we carried out additional $\tau_P$ calculations with 3% vacancies (as in our film). The simulations employ three different stochastic arrangements of point defects. Our classical model shows that N vacancies bind easily to $\{110\}\langle 1\bar{1}0\rangle$ and $\{111\}\langle 1\bar{1}0\rangle$ edge dislocations during their glide. Although Shockley partials move easily in stoichiometric crystal regions, they stop propagating when pinned by an anion vacancy (see supplementary video 2). Accordingly, the Peierls stress required to sustain glide on {111} planes becomes ≈30 times higher ($\tau_P^{\{111\}}$ =6.1±0.1 GPa) than in stoichiometric regions ($\tau_P^{\{111\}}$ ≈0.2 GPa). Our calculations evidence vacancy-pinning effects also for $\{110\}\langle 1\bar{1}0\rangle$ edge dislocations. However, the increase in Peierls stress caused by N vacancies is less pronounced in this case ($\tau_P^{\{110\}}$ = 2.8±0.3 GPa in TiN$_x$ vs $\tau_P^{\{110\}}$ = 0.64±0.03 GPa in TiN). Given that a single vacancy is sufficient to obstruct dislocation motion, we conclude that relevant Peierls stress values for $\{110\}\langle1\bar{1}0\rangle$ and $\{111\}\langle1\bar{1}0\rangle$ edge dislocations are $\tau_P^{\{110\}}$ = 2.8±0.3 GPa and $\tau_P^{\{111\}}$ = 6.1±0.1 GPa. Glide of sessile $\{001\}\langle1\bar{1}0\rangle$ Lomer dislocations, instead, requires very high activation stress ($\tau_P^{\{001\}}$ = 8.4±0.2 GPa) also in stoichiometric TiN.

Our $\tau_P$ results (2.8, 6.1, and 8.4 GPa) are consistent with the Peierls stress assessed for single-crystal TiN during mechanical loading ($\tau_P \approx$ 6.1 GPa, see [35]). The trend in predicted $\tau_P$ values (Table 1) is also in line with experimental observations of preferential $\{110\}\langle1\bar{1}0\rangle$ slip in TiN at room temperature. $\{111\}\langle1\bar{1}0\rangle$ slip may become active at elevated temperatures, similarly to what seen in TiC [62]. In agreement with present observations, anion vacancies are known to enhance hardness in single-crystal TiN [58]. However, it has also been reported that lattice vacancies can have a detrimental role on the hardness of some B1 ceramics [78].

As discussed above, our models suggest that the presence of dislocations locally alters the electronic structure of the host TiN lattice (Figs. 3 and 5). Considering $\{110\}\langle1\bar{1}0\rangle$ edge dislocations as relevant defects, we further investigated their effect on DOS and chemical bonding (Fig. 7). For both defect-free and core regions, the N (s) – Ti (s) states are grouped in a single DOS peak located at a binding energy of ≈15 eV (Fig. 7a). The presence of two secondary peaks approximately 13 and 14 eV below the Fermi energy ($E_F$) is due to the fact that our TiN supercells are periodic only along [001], i.e., terminated by (100) and (010) surfaces (insets in Fig. 7a and Fig. 3a).[f1] The main contribution to nearest-neighbor bonding stems from N (p) – Ti (d) states dispersed in the energy interval between -6 and -2 eV (Fig. 7a). Finally, the DOS of next-nearest-neighbor Ti (d) – Ti (d) bonds ranges from -6 eV up to the Fermi level. These latter states are weakly bonding in defect-free TiN [64]. However, the

---

[f1] The location of surface states in the DOS is determined by comparison with DOS calculations where we use a fully periodic supercell with the same size and accuracy (k-points, energy cutoff). These results are not shown to facilitate readability of Fig. 7. Note also that we employ $\Gamma$-point sampling in present DFT investigations: DOS results exhibit some differences in comparison to calculations with thick k-point meshes.

one-dimensional electron accumulation parallel to the dislocation line seen in Fig. 3e is manifested by a sharp d-d electronic peak close to the Fermi level (see Fig. 7b), which suggests strengthened Ti-Ti bonds around the core.

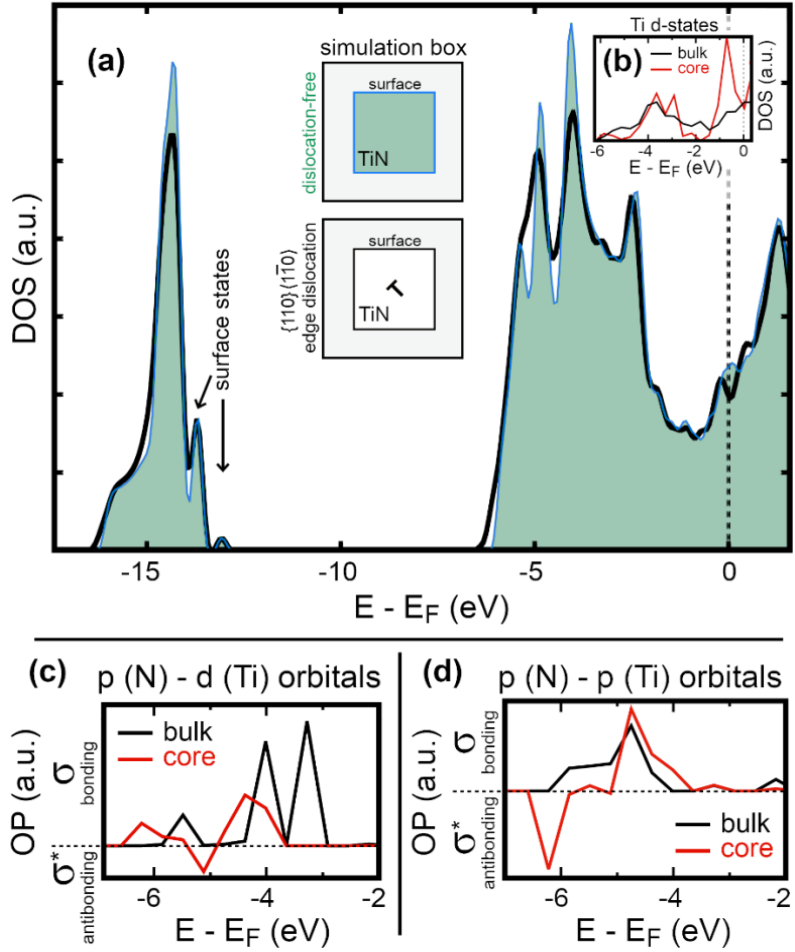

Figure 7. (a,b) Electronic density of states and (c,d) overlap populations calculated for Ti-N interactions in defect-free regions ("bulk") and near the {110}⟨1$\bar{1}$0⟩ edge dislocation ("core"). The DOS in (a) is calculated for a supercell with (black line) and one without the dislocation (green shaded areas). Note that our supercell models are periodic only along one direction (see insets in (a)). Panel (b) illustrates the local electron density of Ti d-states. The DOS is calculated for "bulk" Ti atoms ($Ti_b$) in defect-free domains and for $Ti_1$, $Ti_2$, $Ti'_1$, and $Ti'_2$ atoms near the dislocation core (Ti atoms labelled after Fig. 3).

The global modification of the electronic structure produced by the dislocation core is seen by comparing the DOS of dislocation-free and dislocation-containing supercells (insets of Fig. 7a). The presence of the defect causes an overall broadening and lowering of DOS peaks at binding energies of ≈14.5, 5, and 4 eV, which is indicative of reduced orbital hybridization among nearest neighbors. As expected, the large lattice distortion at the dislocation core weakens Ti–N bonds in relation to defect-free domains. The effect is also reflected by Peierls stress $\tau_P^{\{110\}}$ values (Table 1 and DFT results [23]) that are at least 10 times smaller than the ideal shear strength on the same slip plane ($\gamma^{\{110\}} \approx 30$ GPa [52, 79-81]).

The relative strength of Ti-N bonds in defect-free domains and at the dislocation core can be assessed using DFT-based chemical bonding analyses. The approach – previously implemented for studying bulk defect-free lattices [64] and adatom/surface interactions (see, e.g., Figure 5 in [82] – allows us to evaluate the overlap-population (OP) between atomic orbitals (see Methods section). Specifically, considered $Ti_{core}$–$N_{core}$ pairs are labelled as $N_2$ – $Ti_3$ and $N_1$ – $Ti_7$ in Fig. 3c. The exact location of the glide plane – indicated by a dashed arrow in Fig. 3c – has been revealed by molecular dynamics investigations (see figure 10 in [52]). Our analysis shows that nearest-neighbor σ d (Ti) – p (N) and σ p (Ti) – p (N) orbitals in "bulk" regions (outside red-square area in Fig. 3a) suggest a covalent-bonding character (see Fig. 7c,d), alike defect-free TiN lattices [53, 64]. Conversely, σ d ($Ti_{core}$) – p ($N_{core}$) and p (Ti) – p (N) bonds across the {110} glide plane exhibit lower OP peaks (Fig. 7c), which is signature of reduced bond strength. Moreover, the OP of both σ d ($Ti_{core}$) – p ($N_{core}$) and σ p ($Ti_{core}$) – p ($N_{core}$) orbitals displays σ* antibonding peaks at binding energies of ≈5 and ≈6 eV, respectively (Fig. 7c,d). These antibonding states should additionally favor bond breakage and onset $\{110\}\langle 1\bar{1}0\rangle$ slip during shear deformation.

The observation of partial edge dislocations in as-deposited TiN films (Fig. 4), besides the expected primary $\{110\}\langle 1\bar{1}0\rangle$ edge dislocations (Fig. 2), indicates how TiN crystals may accommodate strains or misfits not considered before. Our experimental findings, corroborated by $\tau_P$ results in Table 1, may also help to rationalize phenomenological observations of unexpected room-temperature plasticity in TiN (see, e.g., [33]). Although it is reasonable to assume that dislocation-mediated plasticity during loading is primarily driven by nucleation and motion of perfect $\{110\}\langle 1\bar{1}0\rangle$ edge dislocations, a modest contribution may also be provided by partial dislocations that glide on {111} planes. Indeed, the $\{111\}\langle 1\bar{1}0\rangle$ slip system provides sufficient degrees of freedom for plastic behavior in cubic crystals [83]. Hence, tuning the synthesis conditions to increase the density of Shockley partials may be used as a dislocation-engineering strategy [4] to toughen typically brittle B1 carbonitrides. At the same time, the introduction of dislocations in the {111}⟨1$\bar{1}$0⟩ slip system has potential implications for enabling strain-hardening mechanisms during load: our microscopy observations indicate that TiN may have the ability to pin dislocations that move on intersecting {111} planes by forming sessile Lomer, or Lomer-Cottrell, junctions (Fig. 6).

The finding of Lomer dislocations is unprecedented for TM nitrides and isostructural carbides. For this reason, the possibility that {100}⟨1$\bar{1}$0⟩ slip may contribute to hardness anisotropy has been traditionally neglected [62]. More recent work, however, phenomenologically evidenced {100}⟨1$\bar{1}$0⟩ slip during compression of bulk single crystal ZrC micropillars [84]. Corroborated by the observations given in Ref. [84], our results suggest that sessile {100}⟨1$\bar{1}$0⟩ edge dislocations may be of general relevance to understand the macroscopic mechanical behavior of B1 nitrides and carbides. Slip on non-close-packed planes should be taken into consideration, especially when changing the load direction. Given that internal stresses in as-deposited films are not sufficient to induce intersection of Shockley partials, our experimental findings (Fig. 6) indicate that {100}⟨1$\bar{1}$0⟩ Lomer junctions are "native" crystallographic defects in TiN. On the other hand, nucleation of {100}⟨1$\bar{1}$0⟩ edge dislocations has also been observed as reaction of {111} stacking faults during simulations of TiN/AlN superlattices subject to strain (see Fig. 3b in Koutna *et al.* [53]).

Depending on the relative density of each line defect, one may expect significant alterations of macroscopic material properties, i.e., mechanical, electrical, and optical properties, which suggests a route for materials design. Note, for example, that a change in the growth direction of single-crystal TiN on MgO has been shown to increase the dislocation density by two orders of magnitude [60]. A recent study [4] also suggests a nanoengineering route to increase the density of mobile dislocations in ceramics. This enhances the material toughness by formation of plastic zones [4]. Furthermore, local lattice distortions and changes in interatomic bond lengths caused by the core may favor trapping of doping elements (impurities) and act as sinks for point-defects (vacancy, interstitials), which further contributes to modify observable properties. As shown by our results, the observed substantial effect of N vacancy-dislocation interaction in Peierls stresses can thus be used as a mean for materials design and can be expanded to the study of effect of impurities and interstitials. Hence, the atomic and electronic structure, bonding character, Peierls stresses, and core energies of different cores merit attention when designing nitride materials at the nanoscale level. We envision that accounting the effect of these findings from atomic-level to large-scale simulations will provide more accurate prediction of a material's macroscopic properties and offer new perspectives for dislocation-engineering. Such defect analyses will also be useful in understanding dislocation effects in other transition-metal nitride films.

## 5. Conclusions

By combination of high-resolution HAADF-STEM imaging, classical simulations, and first principles calculations, we have elucidated a variety of dislocation types in heteroepitaxial magnetron sputter-deposited TiN thin films.

- We found that besides the dislocations on the known primary $\{110\}\langle 1\bar{1}0\rangle$ slip system, dislocations lying on the non-primary slip systems – $\{111\}\langle 1\bar{1}0\rangle$ and $\{001\}\langle 1\bar{1}0\rangle$ – are stored in TiN.
- Classical simulations using MEAM force fields closely model the TiN dislocation core structures which were then analyzed by DFT calculations.
- The DFT-based chemical bonding analyses suggest enhanced metal-metal bonding states at the dislocation cores compared to defect-free regions that locally weakens Ti–N bonds and decreases the directional covalent character.
- Complementary estimates of Peierls stresses of the observed slip systems confirm the preferred slip on $\{110\}\langle 1\bar{1}0\rangle$ but also predict substantial N vacancy-pinning effects of both $\{110\}\langle 1\bar{1}0\rangle$ and $\{111\}\langle 1\bar{1}0\rangle$ dislocations.

The presence of these dislocations and the understanding of their atomic core structures shed light on the role of dislocations in the plastic behavior of transition metal nitrides and have implications on the nanoscale design of hard nitride films. Our results highlight that a variety of dislocations and their core structures should be considered when interpreting and simulating experimental results as well as evaluating the properties of transition metal nitrides.

**Acknowledgments**

Theoretical modelling and calculations were carried out using the resources provided by the Swedish National Infrastructure for Computing (SNIC), partially funded by the Swedish Research Council through Grant Agreement no VR-2015-04630. We acknowledge the financial support given by the following: VINNOVA (FunMat-II project grant no. 2016-05156), the Swedish Research Council (VR grants no 2017-03813, 2017-06701, 2021-04426, and 2021-00357) and grant no 2019-00191 (for accelerator-based ion-technological center in Tandem accelerator laboratory in Uppsala University), and the Swedish government strategic research area grant AFM – SFO MatLiU (2009-00971). IAA and FT acknowledge support from the Knut and Alice Wallenberg Foundation (Wallenberg Scholar grant no. KAW-2018.0194).

**REFERENCES**

[1] D. Hull, D.J. Bacon, Introduction to Dislocations, Butterworth-Heinemann, Oxford, 2011.
[2] J.P. Hirth, J. Lothe, Theory of Dislocations, McGraw-Hill, New York, 1968.
[3] N. Shibata, M.F. Chisholm, A. Nakamura, S.J. Pennycook, T. Yamamoto, Y. Ikuhara, Nonstoichiometric Dislocation Cores in α-Alumina, Science 316(5821) (2007) 82-85.
[4] L. Porz, A.J. Klomp, X. Fang, N. Li, C. Yildirim, C. Detlefs, E. Bruder, M. Hoefling, W. Rheinheimer, E.A. Patterson, P. Gao, K. Durst, A. Nakamura, K. Albe, H. Simons, J. Roedel, Dislocation-toughened ceramics, MATERIALS HORIZONS 8(5) (2021) 1528-1537.
[5] X. Zhou, J.R. Mianroodi, A.K. Da Silva, T. Koenig, G.B. Thompson, P. Shanthraj, D. Ponge, B. Gault, B. Svendsen, D. Raabe, The hidden structure dependence of the chemical life of dislocations, Science Advances 7(16) (2021).
[6] I. Arslan, N.D. Browning, Intrinsic electronic structure of threading dislocations in GaN, Physical Review B - Condensed Matter and Materials Physics 65(7) (2002) 0753101-07531010.
[7] Z. Wang, M. Saito, Y. Ikuhara, K.P. McKenna, Polymorphism of dislocation core structures at the atomic scale, Nature Communications 5 (2014).
[8] B. Feng, R. Ishikawa, A. Kumamoto, N. Shibata, Y. Ikuhara, Atomic Scale Origin of Enhanced Ionic Conductivity at Crystal Defects, Nano Letters 19(3) (2019) 2162-2168-2168.
[9] I. Sugiyama, N. Shibata, Z. Wang, S. Kobayashi, T. Yamamoto, Y. Ikuhara, Ferromagnetic dislocations in antiferromagnetic NiO, Nature Nanotechnology 8(4) (2013) 266-270.
[10] H. Hojo, E. Tochigi, N. Shibata, B. Feng, Y. Ikuhara, T. Mizoguchi, H. Ohta, Atomic structure and strain field of threading dislocations in CeO 2 thin films on yttria-stabilized ZrO2, Applied Physics Letters 98(15) (2011).
[11] Y. Qian, Q. Liang, T. Tomohito, T. Rachel, R. David, J. J. W. Morris, A. Mark, D.C. Chrzan, M.M. Andrew, Origin of dramatic oxygen solute strengthening effect in titanium, Science 347(6222) (2015) 635-639.
[12] B. Yang, Y.T. Zhou, D. Chen, X.L. Ma, Local decomposition induced by dislocation motions inside precipitates in an Al-alloy, Scientific Reports 3 (2013).
[13] M. Garbrecht, J.L. Schroeder, L. Hultman, B. Saha, T.D. Sands, Dislocation-pipe diffusion in nitride superlattices observed in direct atomic resolution, Scientific Reports 7 (2017).
[14] F. Zhang, A. Walker, K. Wright, J. Gale, Defects and dislocations in MgO: atomic scale models of impurity segregation and fast pipe diffusion, JOURNAL OF MATERIALS CHEMISTRY 20(46) (2010) 10445-10451.
[15] H. Wu, J. Leng, X. Teng, Q. Li, J. Li, J. Wu, D. Xu, T. Su, Y. Zhu, Characterizing the interactions of edge dislocation dipole in hexagonal close packed Ti-Al alloys, Materials and Design 164 (2019).
[16] N. Li, N. Mara, S.K. Yadav, X.Y. Liu, R.G. Hoagland, J. Wang, A. Misra, Quantification of dislocation nucleation stress in TiN through high-resolution in situ indentation experiments and first principles calculations, Scientific Reports 5 (2015).
[17] P.H. Mayrhofer, C. Mitterer, L. Hultman, H. Clemens, Microstructural design of hard coatings, Progress in Materials Science 51(8) (2006) 1032-1114.
[18] J. Buchinger, A. Wagner, P.H. Mayrhofer, M. Bartosik, L. Löfler, D. Holec, J. Ast, J. Michler, Z. Chen, Z.L. Zhang, Fracture properties of thin film TiN at elevated temperatures, Materials and Design 194 (2020).

[19] Z. Zhang, A. Ghasemi, N. Koutná, Z. Xu, T. Grünstäudl, K. Song, D. Holec, Y. He, P.H. Mayrhofer, M. Bartosik, Correlating point defects with mechanical properties in nanocrystalline TiN thin films, Materials & Design 207 (2021).
[20] F. Magnus, A.S. Ingason, S. Olafsson, J.T. Gudmundsson, Growth and in-situ electrical characterization of ultrathin epitaxial TiN films on MgO, Thin Solid Films 519(18) (2011) 5861-5867.
[21] M. Ragini, C. Ching-Wen, D. Abhishek, C. Zong-Yi, Y. Ta-Jen, L. Ho Wai Howard, L. Yu-Jung, G. Shangjr, Optimized Titanium Nitride Epitaxial Film for Refractory Plasmonics and Solar Energy Harvesting, American Chemical Society (ACS), 2021, pp. 13658-13665.
[22] T. Krekeler, S.S. Rout, G.V. Krishnamurthy, M. Störmer, M. Arya, A. Ganguly, D.S. Sutherland, S.I. Bozhevolnyi, M. Ritter, K. Pedersen, A.Y. Petrov, M. Eich, M. Chirumamilla, Unprecedented Thermal Stability of Plasmonic Titanium Nitride Films up to 1400 °C, Advanced Optical Materials 9(16) (2021) 1-11.
[23] S.K. Yadav, X.Y. Liu, R. Ramprasad, A. Misra, Core structure and Peierls stress of edge and screw dislocations in TiN: A density functional theory study, Acta Materialia 74 (2014) 268-277.
[24] M. Odén, H. Ljungcrantz, L. Hultman, Characterization of the induced plastic zone in a single crystal TiN(001) film by nanoindentation and transmission electron microscopy, Journal of Materials Research 12(8) (1997) 2134-2142.
[25] S.K. Yadav, X.Y. Liu, J. Wang, R.G. Hoagland, R. Ramprasad, A. Misra, First-principles density functional theory study of generalized stacking faults in TiN and MgO, Philosophical Magazine 94(5) (2014) 464-475.
[26] W. Cai, V.V. Bulatov, J. Chang, J. Li, S. Yip, Dislocation Core Effects on Mobility, in: J.P. Hirth (Ed.), Dislocations in Solids, Elsevier B.V.2004.
[27] N. De Leon, X.X. Yu, G.B. Thompson, H. Yu, C.R. Weinberger, Bonding effects on the slip differences in the B1 monocarbides, Physical Review Letters 114(16) (2015).
[28] R.F. Zhang, S.H. Sheng, S. Veprek, Origin of different plastic resistance of transition metal nitrides and carbides: Stiffer yet softer, Scripta Materialia 68(12) (2013) 913-916.
[29] D. Rodney, L. Ventelon, E. Clouet, L. Pizzagalli, F. Willaime, Ab initio modeling of dislocation core properties in metals and semiconductors, Acta Materialia 124 (2017) 633-659.
[30] E. Maras, M. Saito, K. Inoue, H. Jónsson, Y. Ikuhara, K.P. McKenna, Determination of the structure and properties of an edge dislocation in rutile TiO2, Acta Materialia 163 (2019) 199-207.
[31] L.J. Teutonico, The dissociation of dislocations in the face-centered cubic structure, Acta Metallurgica 11(12) (1963) pp. 1283–1289.
[32] A. Minor, E. Stach, J. Morris, I. Petrov, In-situ nanoindentation of epitaxial TiN/MgO (001) in a transmission electron microscope, Journal of Electronic Materials 32(10) (2003) 1023.
[33] S. Kumar, D.E. Wolfe, M.A. Haque, Dislocation shielding and flaw tolerance in titanium nitride, International Journal of Plasticity 27(5) (2011) 739-747.
[34] N. Li, H. Wang, A. Misra, J. Wang, In situ Nanoindentation Study of Plastic Co-deformation in Al-TiN Nanocomposites, Scientific Reports (2014) 1-6.
[35] F. Giuliani, C. Ciurea, V. Bhakhri, M. Werchota, L.J. Vandeperre, P.H. Mayrhofer, Deformation behaviour of TiN and Ti–Al–N coatings at 295 to 573 K, Thin Solid Films 688 (2019).

[36] Z. Xu, Z. Zhang, M. Bartosik, Y. Zhang, P.H. Mayrhofer, Y. He, Insight into the structural evolution during TiN film growth via atomic resolution TEM, Journal of Alloys and Compounds 754 (2018) 257-267.

[37] D.G. Sangiovanni, L. Hultman, V. Chirita, I. Petrov, J.E. Greene, Effects of phase stability, lattice ordering, and electron density on plastic deformation in cubic TiWN pseudobinary transition-metal nitride alloys, Acta Materialia 103 (2016) 823-835.

[38] H. Kindlund, D.G. Sangiovanni, I. Petrov, J.E. Greene, L. Hultman, A review of the intrinsic ductility and toughness of hard transition-metal nitride alloy thin films, Thin Solid Films 688 (2019).

[39] K.M. Calamba, J. Barrirero, R. Boyd, M.P. Johansson Jõesaar, M. Odén, J.F. Pierson, S. Bruyère, A.L. Febvrier, P. Eklund, F. Mücklich, Dislocation structure and microstrain evolution during spinodal decomposition of reactive magnetron sputtered heteroepixatial c-(Ti0.37,Al0.63)N/c-TiN films grown on MgO(001) and (111) substrates, Journal of Applied Physics 125(10) (2019).

[40] K.M. Calamba, J. Salamania, M.P.J. Jõesaar, L.J.S. Johnson, R. Boyd, J.F. Pierson, M.A. Sortica, D. Primetzhofer, M. Odén, Effect of nitrogen vacancies on the growth, dislocation structure, and decomposition of single crystal epitaxial (Ti1-xAlx)Ny thin films, Acta Materialia 203 (2021).

[41] J. Salamania, L.J.S. Johnson, I.C. Schramm, K.M. Calamba, R. Boyd, B. Bakhit, L. Rogström, M. Odén, Influence of pulsed-substrate bias duty cycle on the microstructure and defects of cathodic arc-deposited Ti1-xAlxN coatings, Surface & Coatings Technology 419 (2021).

[42] J.M. Molina-Aldareguia, S.J. Lloyd, M. Odén, T. Joelsson, L. Hultman, W.J. Clegg, Deformation structures under indentations in TiN/NbN single-crystal multilayers deposited by magnetron sputtering at different bombarding ion energies, Philosophical Magazine A 82(10) (2002) 1983-1992.

[43] C. Sun, G. Lian, J. Wang, M.J. Kim, T. Paulauskas, C. Buurma, R.F. Klie, F.G. Sen, M.K.Y. Chan, Atomic and electronic structure of Lomer dislocations at CdTe bicrystal interface, Scientific Reports 6 (2016).

[44] A. Le Febvrier, J. Jensen, P. Eklund, Wet-cleaning of MgO (001): Modification of surface chemistry and effects on thin film growth investigated by x-ray photoelectron spectroscopy and time-of-flight secondary ion mass spectroscopy, Journal of Vacuum Science & Technology A: Vacuum, Surfaces, and Films 35(2) (2017) 021407.

[45] B. Bakhit, D. Primetzhofer, E. Pitthan, M.A. Sortica, E. Ntemou, J. Rosen, L. Hultman, I. Petrov, G. Greczynski, Systematic compositional analysis of sputter-deposited boron-containing thin films, JOURNAL OF VACUUM SCIENCE & TECHNOLOGY A 39(6) (2021) 063408.

[46] R. Lin, R. Zhang, C. Wang, X.-Q. Yang, H.L. Xin, TEMImageNet training library and AtomSegNet deep-learning models for high-precision atom segmentation, localization, denoising, and deblurring of atomic-resolution images, Scientific Reports 11(1) (2021) 1-15.

[47] J. Barthel, Dr. Probe: A software for high-resolution STEM image simulation, Ultramicroscopy 193 (2018) 1-11.

[48] G.A. Almyras, D.G. Sangiovanni, K. Sarakinos, Semi-Empirical Force-Field Model for the Ti1–xAlxN (0 ≤ x ≤ 1) System, Materials, Vol 12, Iss 2, p 215 (2019 (2019).

[49] P. Hirel, Atomsk: A tool for manipulating and converting atomic data files, Computer Physics Communications 197 (2015) 212-219.

[50] S. Plimpton, Fast parallel algorithms for short-range molecular dynamics, Journal of Computational Physics 117(1) (1995) 1-19.
[51] Z. Chen, Y. Zheng, L. Loefler, M. Bartosik, G.K. Nayak, O. Renk, D. Holec, P.H. Mayrhofer, Z. Zhang, Atomic insights on intermixing of nanoscale nitride multilayer triggered by nanoindentation, Acta Materialia 214 (2021).
[52] D.G. Sangiovanni, A. Kraych, M. Mrovec, J. Salamania, M. Oden, F. Tasnadi, I.A. Abrikosov, Descriptors for mechanical strength and slip-induced crack-blunting in refractory ceramics, arXiv preprint arXiv:2203.00622 (2022).
[53] N. Koutná, L. Löfler, D. Holec, Z. Chen, Z. Zhang, L. Hultman, P.H. Mayrhofer, D.G. Sangiovanni, Atomistic mechanisms underlying plasticity and crack growth in ceramics: a case study of AlN/TiN superlattices, Acta Materialia 229 (2022).
[54] G. Kresse, From ultrasoft pseudopotentials to the projector augmented-wave method, Physical Review B - Condensed Matter and Materials Physics 59(3) (1999) 1758-1775.
[55] G. Kresse, J. Furthmueller, Efficiency of ab-initio total energy calculations for metals and semiconductors using a plane-wave basis set, COMPUTATIONAL MATERIALS SCIENCE 6(1) (1996) 15-50.
[56] J.P. Perdew, K. Burke, M. Ernzerhof, Generalized gradient approximation made simple, Physical Review Letters 77(18) (1996) 3865-3868.
[57] R. Hoffmann, Solids and Surfaces : A Chemist's View of Bonding in Extended Structures, Wiley-VCH, Ithaca, New York, 1988.
[58] C.S. Shin, D. Gall, N. Hellgren, J. Patscheider, I. Petrov, J.E. Greene, Vacancy hardening in single-crystal $TiN_x(001)$ layers, 2003, pp. 6025-6028.
[59] A. Kelly, P. Gay, P.B. Hirsch, The estimation of dislocation densities in metals from X-ray data, Acta Metallurgica 1(3) (1953) 315-319.
[60] H. Ljungcrantz, M. Odén, L. Hultman, J.E. Greene, J.E. Sundgren, Nanoindentation studies of single-crystal (001)-, (011)-, and (111)-oriented TiN layers on MgO, Journal of Applied Physics 80(12) (1996) 6725.
[61] V. Grillo, The effect of surface strain relaxation on HAADF imaging, Ultramicroscopy 109(12) (2009) 1453-1464.
[62] R. H. J. Hannink, D. L. Kohlstedt, M. J. Murray, Slip System Determination in Cubic Carbides by Hardness Anisotropy, Proceedings of the Royal Society of London. Series A, Mathematical and Physical Sciences 326(1566) (1972) 409-420.
[63] D.G. Sangiovanni, Inherent toughness and fracture mechanisms of refractory transition-metal nitrides via density-functional molecular dynamics, Acta Materialia 151 (2018) 11-20.
[64] D.G. Sangiovanni, L. Hultman, V. Chirita, Supertoughening in B1 transition metal nitride alloys by increased valence electron concentration, Acta Materialia 59 (2011) 2121.
[65] A.B. Mei, H. Kindlund, E. Broitman, L. Hultman, I. Petrov, J.E. Greene, D.G. Sangiovanni, Adaptive hard and tough mechanical response in single-crystal B1 $VN_x$ ceramics via control of anion vacancies, Acta Materialia (2020).
[66] H. Yu, M. Bahadori, C.R. Weinberger, G.B. Thompson, Understanding dislocation slip in stoichiometric rocksalt transition metal carbides and nitrides, Journal of Materials Science 52(11) (2017) 6235-6248.
[67] L. Wang, X. Han, P. Liu, Y. Yue, Z. Zhang, E. Ma, In Situ observation of dislocation behavior in nanometer grains, Physical Review Letters 105(13) (2010).
[68] C. Frigeri, S. Bietti, A. Scaccabarozzi, R. Bergamaschini, C.V. Falub, V. Grillo, M. Bollani, E. Bonera, P. Niedermann, H. von Kanel, A Structural Characterization of GaAs MBE Grown on Si Pillars, ACTA PHYSICA POLONICA SERIES A 125(4) (2014) 986-990.

[69] M.S. Colla, H. Idrissi, T. Pardoen, B. Amin-Ahmadi, D. Schryvers, L. Malet, S. Godet, J.P. Raskin, Dislocation-mediated relaxation in nanograined columnar palladium films revealed by on-chip time-resolved HRTEM testing, Nature Communications 6 (2015).
[70] D. Rodney, R. Phillips, Structure and strength of dislocation junctions: An atomic level analysis, Physical Review Letters 82(8) (1999) 1704-1707.
[71] X.D. Xu, P. Liu, Z. Tang, A. Hirata, S.X. Song, T.G. Nieh, P.K. Liaw, C.T. Liu, M.W. Chen, Transmission electron microscopy characterization of dislocation structure in a face-centered cubic high-entropy alloy $Al_{0.1}CoCrFeNi$, Acta Materialia 144 (2018) 107-115.
[72] A.F. Marshall, D.B. Aubertine, W.D. Nix, P.C. McIntyre, Misfit dislocation dissociation and Lomer formation in low mismatch SiGe/Si heterostructures, Journal of Materials Research 20 (2005) pp. 447–455.
[73] T.C. Lin, Y.C. Huang, C.C. Yen, C.H. Tung, Y.T. Hsiao, S.Y. Chang, S.Y. Lin, Small-Size-Induced Plasticity and Dislocation Activities on Non-Charge-Balanced Slip System of Ionic MgO Pillars, Nano Letters 18(8) (2018) 4993-5000.
[74] P. Haasen, Dislocations and the plasticity of ionic crystals, Materials Science and Technology 1(12) (1985) 1013-1024.
[75] G.W. Groves, A. Kelly, Independent slip systems in crystals, Philosophical Magazine 8(89) (1963) 877-887.
[76] D.J. Rowcliffe, Hollox, G.E., Hardness anisotropy, deformation mechanisms and brittle-to-ductile transition in carbide, *J Mater Sci* 6 (1971) 1270–1276.
[77] X.W. Zhou, R.B. Sills, D.K. Ward, R.A. Karnesky, Atomistic calculations of dislocation core energy in aluminium, Physical Review B 95(5) (2017).
[78] C. Hu, X. Zhang, Z. Gu, H. Huang, S. Zhang, X. Fan, W. Zhang, Q. Wei, W. Zheng, Negative effect of vacancies on cubic symmetry, hardness and conductivity in hafnium nitride films, Scripta Materialia 108 (2015) 141-146.
[79] S.H. Jhi, S.G. Louie, M.L. Cohen, J. Ihm, Vacancy Hardening and Softening in Transition Metal Carbides and Nitrides, AMERICAN PHYSICAL SOCIETY, United States, 2001, p. 3348.
[80] S.H. Jhi, S.G. Louie, M.L. Cohen, J.W. Morris, Mechanical instability and ideal shear strength of transition metal carbides and nitrides, Physical Review Letters 87(7) (2001) 75503-1-75503-4.
[81] R.F. Zhang, S.H. Sheng, S. Veprek, First principles studies of ideal strength and bonding nature of AlN polymorphs in comparison to TiN, Applied Physics Letters 91(3) (2007).
[82] D.G. Sangiovanni, L. Hultman, V. Chirita, I. Petrov, J.E. Greene, A.B. Mei, Ab Initio Molecular Dynamics Simulations of Nitrogen/VN(001) Surface Reactions: Vacancy-Catalyzed $N_2$ Dissociative Chemisorption, N Adatom Migration, and $N_2$ Desorption, Journal of Physical Chemistry C 120(23) (2016) 12503-12516.
[83] R. von Mises, Mechanik der plastischen formänderung von kristallen, Z. Angew. Math. Mech. 8(161) (1928).
[84] S. Kiani, C. Ratsch, A.M. Minor, S. Kodambaka, J.M. Yang, Orientation- and size-dependent room-temperature plasticity in ZrC crystals, Philosophical Magazine 95 (2015) 985-997.